\documentclass[twocolumn,aps,showpacs,prb,tightenlines,amsmath,amssymb]{revtex4}
\usepackage[dvips]{graphicx}
\usepackage{dcolumn}
\usepackage{bm}
\usepackage{colordvi}

\begin{document}

\newcommand{\bgreek}[1]{\mbox{\boldmath$#1$\unboldmath}}

\title{Kinetic investigation on extrinsic spin Hall effect
 induced by skew scattering}
\author{J. L. Cheng}
\author{M. W. Wu}
\thanks{Author to whom all correspondence should be addressed}
\email{mwwu@ustc.edu.cn.}
\affiliation{Hefei National Laboratory for Physical Sciences at
Microscale and Department of Physics, University of Science and Technology of China, Hefei,
Anhui, 230026, China}
\date{\today}

\begin{abstract}
The kinetics of the extrinsic spin Hall conductivity induced by the
skew scattering is performed from the fully microscopic kinetic
spin Bloch equation approach in $(001)$ GaAs symmetric quantum well. 
In the steady state, 
the extrinsic spin Hall current/conductivity 
vanishes for the linear-$\mathbf k$ dependent spin-orbit coupling
and is very small for the
cubic-$\mathbf k$ dependent spin-orbit coupling. 
The spin precession induced by the Dresselhaus/Rashba
spin-orbit coupling plays a very important role in the vanishment of
the extrinsic spin Hall conductivity in the steady state.  An in-plane
spin polarization is induced by the skew scattering, with the help of
the spin-orbit coupling. This spin polarization is very different from
the current-induced spin polarization.

\end{abstract}

\maketitle

Generating and manipulating the spin polarization in semiconductors is
one of the important prerequisites for the realization of the new spintronic
device.\cite{spintronics}
Spin Hall effect (SHE) is considered as the
convenient method to generate spin polarization
in additional to the traditional methods such as the
external magnetic field, the circular/linear
polarized laser,\cite{meier,tarasenko05} the
spin-galvanic effect\cite{ganlchev} and the spin injection from the
ferromagnetic metal to semiconductor.\cite{ohno}
The SHE is induced by intrinsic
or extrinsic spin-orbit coupling (SOC)\cite{hans06}
which gives rise to the spin current vertical to the charge current
without applying external magnetic field and/or spin
accumulation at sample edges.\cite{zsf,tse05,nomura}
Experimentally, the spin Hall conductivity (SHC) is estimated
indirectly by the spin accumulation at the sample
edges,\cite{kato.sc04,sih.na05,sih.prl06,stern.prl06,wunderlich.prl05}
or the charge current with a transverse magnetic field applied in
a gyrotropic systems.\cite{ganichev.prl01,ganichev.na06,ganichev.cn06}
Recently, a direct electronic measurements
of the SHE is given by Valenzuela and Tinkham in the metallic
conductor. \cite{valenzuela.na06}
All these effects are explained as the intrinsic
\cite{murakami03,sinova04,hans06,schliemann06,mishchenko,lsy06a,chen05,sheng05,alexander,shytov} and/or the extrinsic
SHE\cite{zsf,dp.ss,hirsch,tse06-a,tse06-b,hans06,sipe,kavokin,lsy06,huang,hans05,hankiewicz,roksana} theoretically by using the
Kubo formula\cite{murakami03,sinova04,tse06-a,tse06-b} or the Boltzmann
equation\cite{zsf,lsy06,lsy06a,hans06,hans05} with only the
carrier-impurity scattering included.

The intrinsic SHE is induced by the intrinsic SOC
(i.e., the Dresselhaus\cite{Dress} and/or the Rashba\cite{Rashba}
 SOCs) with the applied
external electric field and the resulting SHC was at first
thought as dissipationless one
in perfect crystals.\cite{murakami03,sinova04}
Later investigations proved that this kind of SHC
disappears even for the infinitesimal impurity density with the vertex
correction\cite{lsy06a,inoue,mishchenko,murakami04,chalaev,dimitrova} in the
Rashba model or the linear Dresselhaus model in quantum
wells,\cite{lsy0502392} but remains a finite value for the cubic
SOC.\cite{bernevig0412550} However, the spin current is not an
observable quantity. The Laughlin's gauge gedanken experiment
indicates that the intrinsic SHE cannot lead to any spin
accumulation at sample edges,\cite{sheng05,chen05} unless in
a mesoscopic system.
In additional to the Dresselhaus and the Rashba SOC, the
mixing between the valance and the conduction bands  gives rise to two
corrections: One is the
additional spin-dependent electron-impurity\cite{meier,averkiev} or
the electron-phonon\cite{ganichev.na06,ganichev.cn06} skew
scattering. The extrinsic SHE induced by the skew scattering
  alone has been widely studied by using both the Kubo formula 
and the Boltzmann equation
  method,\cite{zsf,dp.ss,hirsch,tse06-a,hans06,lsy06,huang,hans05}
 and the nonzero extrinsic SHC is obtained. Lately, Tse and Das
  Sarma\cite{tse06-b} have proved the vanishment of
  the extrinsic SHC by considering the vortex correction of the linear
SOC in the Kubo formula. However, a fully microscopic 
calculation of the extrinsic SHC from the kinetic equation approach
 is still missing and the vanishment of the  extrinsic SHC
by the vortex correction from the Kubo approach needs to be verified 
from the kinetic approach.
The other is the additional spin-dependent position and
velocity operators which bring the correction to the definition of the
spin current, and are referred to as
the side-jump mechanism.\cite{hans06} Here
two corrections on the definition should be specified. The
first comes
from the electrical potential which gives an intrinsic-like
contribution and again cannot contribute to the spin accumulation
according to Refs.\ \onlinecite{sheng05,chen05,weng}.  The second
comes from the spin dependent scattering in which
high order correlations between different wave
vectors need to be considered.\cite{lsy06,hans06} It is hard to include
the second correction in the Boltzmann equation approach,\cite{hans06}
though it also gives rise to the
extrinsic SHE, together with the skew scattering. 
In the following, we concentrate
 on the extrinsic SHE induced by the skew scattering.

The spin polarization is not only accumulated at the sample
edges due to the extrinsic SHE, but also observed simultaneously
inside the samples which is induced by the charge
current.\cite{kato.na03,kato.sc04,sih.na05,stern.prl06,hans07,tarasenko06,trushin} By comparing the experiments in
Refs.\ \onlinecite{kato.na03,kato.sc04,stern.prl06}, it is
easy to find that both
spin polarizations are in the same order. In theory,
Engel {\it et al.}\cite{hans07} and Trushin and Schliemann\cite{trushin}
attributed the current-induced spin polarization (CISP)
in the homogeneous system as the results of the current induced
effective magnetic field (EMF) from the SOC.\cite{kato.na03}
Tarasenko showed that the spin-flip phonon
scattering in asymmetric quantum wells can also induce this kind of
spin polarization.\cite{tarasenko06}
However, the spin-conserving skew scattering in two dimensional GaAs
semiconductor can also induce spin polarization inside the sample due
to the extrinsic SHE. This effect has not been studied in the literature.
Moreover, a fully microscopic kinetic investigation on the extrinsic SHE is
also missing in the literature. 

In this paper, we focus on the kinetic process of the extrinsic SHE
induced by the skew scattering  in symmetric  GaAs $(001)$  quantum well
from the kinetic spin  Bloch equation
(KSBE) approach.\cite{wu,wu1} We demonstrate
the important role of the spin precession induced by the (intrinsic)
Dresselhaus/Rashba
SOC to the SHE and show that it is inadequate to study the extrinsic
SHE from the Kubo formalism without considering the
Dresselhaus/Rashba SOC in the literature.
We further show that the extrinsic SHE can generate
spin polarizations in homogeneous system.

By using the non-equilibrium Green function method and the generalized
Kadanoff-Baym Ansatz,\cite{haug} we construct the KSBE\cite{wu,wu1} for
electrons as follows
\begin{equation}
\frac{\partial \rho_{\mathbf k}(t)}{\partial t} - eE\frac{\partial
  \rho_{\mathbf k}}{\partial k_x} + \left.\frac{\partial \rho_{\mathbf
      k}}{\partial t}\right|_{coh}  + \left.
\frac{\partial \rho_{\mathbf k}}{\partial
    t}\right|_{scat} + \left.\frac{\partial \rho_{\mathbf
      k}}{\partial t}\right|_{ss} =0\ .
\label{eq:ksbe}
\end{equation}
Here, $\rho_{\mathbf k}(t)=\begin{pmatrix} f_{\mathbf
    k\uparrow}&\rho_{\mathbf k\uparrow\downarrow}\\\rho_{\mathbf
    k\downarrow\uparrow} &f_{\mathbf k\downarrow}\end{pmatrix}$ is
electron density matrix with wave vector $\mathbf k$ at time
$t$. The applied electric field $\mathbf E$ is assumed
along the $x$-axis  and the magnetic
 field $\mathbf B$ is along the $x$-$y$ (well) plane. The
coherent term describes the spin precession along the
magnetic/effective magnetic field and 
is given by $\left.\frac{\partial \rho_{\mathbf
      k}}{\partial t}\right|_{coh}=i
[\frac{1}{2}({\bf \Omega}^D(\mathbf k)+g\mu_B\mathbf
B)\cdot\bgreek{\sigma}, \rho_{\mathbf k}(t)]$,
where  ${\bf \Omega}^D(\mathbf k)=\gamma
(k_x(k_y^2-(\frac{\pi}{a})^2), k_y((\frac{\pi}{a})^2-k_x^2),0)$ 
represents the EMF from the
Dresselhaus SOC\cite{Dress} with $\gamma$ standing for
the material-determined  SOC strength\cite{gamma,zhou} and $a$
being the well width. Infinite-well-depth assumption is adopted here
and only the lowest subband is taken into account due to small well width.
The spin conserving scattering
$\left.\frac{\partial \rho_{\mathbf k}}{\partial
  t}\right|_{scat}$ is given by the spin conserving electron-impurity
scattering, the electron-phonon scattering and the
electron-electron Coulomb scattering which are given in detail in
Ref.\ \onlinecite{mq}.

\begin{figure}[htp]
\begin{center}\includegraphics[height=5.cm]{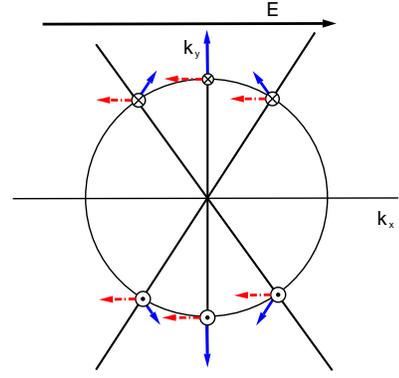}\end{center}
\caption{(Color online) The schematic for the
skew scattering and the intrinsic spin
  orbit coupling in the $\mathbf k$-space. $\odot$ ($\otimes$) stands
 for the spin-up (-down) polarization. 
The blue arrows give the direction of the EMF 
from the Dresselhaus SOC which rotates the spin
  polarization to the direction indicated by the red dashed arrows.}
\label{fig:scheme}
\end{figure}

The skew scattering  is given by the third
order expansion of the electron-impurity scattering and
reads:\cite{hans06,tse06-a,tse06-b}
\begin{eqnarray}
&&\left.\frac{\partial
    \rho_{\mathbf
      k}}{\partial
    t}\right|_{ss}=-2\pi^2N_i\lambda\gamma^2\sum_{\mathbf k_1\mathbf
  k_2;q_1q_2}\delta(\varepsilon_{\mathbf
  k_2}-\varepsilon_{\mathbf k})\delta(\varepsilon_{\mathbf
  k_1}-\varepsilon_{\mathbf k})\nonumber\\
&&\times V(\mathbf k-\mathbf
k_1,q_1)V(\mathbf k_1-\mathbf
k_2,q_2)V(\mathbf k_2-\mathbf
k,-q_1-q_2)\nonumber\\
&&\times\{(\mathbf k-\mathbf k_1)\times(\mathbf
k_2-\mathbf k)\cdot\mbox{\boldmath$\sigma$\unboldmath},\rho_{\mathbf
  k_2}\}\ .
\label{eq:ss}
\end{eqnarray}
Here  $\lambda\gamma^2=\frac{\eta(2-\eta)}{2m^{\ast}E_g(3-\eta)}$ is the
  strength of the skew scattering with
  $\eta=\frac{\Delta}{\Delta+E_g}$.\cite{meier} $E_g$ and $\Delta$ are
 the band gap and the spin-orbit splitting of the valance band
  separately. $V(\mathbf
q,q_z)=\frac{-e^2}{\mathbf q^2+q_z^2+\kappa^2}I(\frac{q_za}{2\pi})$
with $I(x)=e^{i\pi x}\frac{\sin(\pi x)}{\pi
  x(1-x^2)}$ being the form factor and  $\kappa$
denoting the Debye-H\"ucke screening constant.
This term produces an asymmetric spin-conserving
 scattering of electrons so that the
spin-up/down electrons prefer to be scattered to the left/right side.
A schematic for the skew scattering is
given in Fig.\ \ref{fig:scheme} for a left moving electron.

We first analysis the KSBE for some simple cases. After carrying out
the summation over ${\bf k}$, one gets the equation of continuity for the
spin density $\mathbf S = \sum_{\mathbf k}s_{\mathbf
k}=\sum_{\mathbf
  k}\mbox{Tr}[\rho_{\mathbf k}\bgreek{\sigma}]$ as
\begin{equation}
\frac{\partial \mathbf S(t)}{\partial t} - \sum_{\mathbf
  k}{\bf \Omega}^D(\mathbf k)\times\mathbf s_{\mathbf k}(t)
- g\mu_B\mathbf B\times \mathbf S(t) = 0\ .
\label{eq:sj}
\end{equation}
The summation of the spin-conserving scattering is naturally zero.
The summation of the skew-scattering
 in Eq.\ (\ref{eq:ss}) for the two-dimension system is also zero 
because it can only skew the spin-up and -down electrons 
separately instead of flip them. Equation\ (\ref{eq:sj}) is
consistent with the result in Ref.\ \onlinecite{dimitrova} which is
obtained by using the general operator commutation relations.
When only the linear-${\bf k}$
terms in ${\bf \Omega}^D(\mathbf k)$ is retained,
for the steady state one obtains
\begin{equation}
J_y^z=\frac{e g\mu_B B_y}{\gamma m^{\ast}((\pi/a)^2-\langle
  k_x^2\rangle)}S^z\ .
\label{eq:sjd}
\end{equation}
Here $\mathbf{J}^z=-e\sum_{\mathbf k}\hbar 
\mathbf{k}/m^{\ast}(f_{\mathbf k\uparrow}-f_{\mathbf
  k\downarrow})$ is the spin current which is the simplest but the most
widely-used definition\cite{hans06} without considering 
contribution from the  side-jump
effect. The  SHC is given as $\sigma_y^z=J_y^z/E$.
In the derivation, we also take the
approximation $\sum_{\mathbf k}k_xk_y^2s^z_{\mathbf
  k}=-\frac{m^{\ast}}{e}\langle
k_y^2\rangle J_x^z$ and $\sum_{\mathbf   k}k_yk_x^2s^z_{\mathbf
  k}=-\frac{m^{\ast}}{e}\langle k_x^2\rangle J_y^z$ with 
$\langle k_{x/y}^2\rangle$, the average 
value of $k_{x/y}^2$. If the strain-induced SOC $H_{s}=\alpha(k_y\sigma_x -
k_x\sigma_y)$ (with the same expression of the Rashba SOC)
is taken into account, Eq.\ (\ref{eq:sjd}) changes into
\begin{equation}
J_y^z=\frac{\gamma((\frac{\pi}{a})^2-\langle k_y^2\rangle)B_y-2\alpha
  B_x}{\gamma^2((\frac{\pi}{a})^2-\langle k_x^2\rangle)(
(\frac{\pi}{a})^2-\langle
  k_y^2\rangle) -4\alpha^2}\frac{eg\mu_B}{m^{\ast}}S^z\ .
\label{eq:sjdr}
\end{equation}
Without the external magnetic field,  Eqs.\ (\ref{eq:sjd}) and 
(\ref{eq:sjdr}) give
$J_y^z=0$, which verifies the zero extrinsic SHC given by Ref.\
  [\onlinecite{tse06-b}] from the Kubo approach.
However, the skew
scattering does generate the spin currents when the electric field is
applied (in Fig.\ \ref{fig:scheme}), but it is just the dynamic
one. To make clear how the spin current disappears, we
rewrite the $x$-component of Eq.\ (\ref{eq:sj}) as
\begin{equation}
{\partial S^x}/{\partial
  t}=m^{\ast}\gamma[(\pi/a)^2-\langle k_z^2\rangle]J_y^z/e\ .
\label{eq:sxjyz}
\end{equation}
It is hence easy to find that the spin-Hall current is converted to the 
spin polarization along the $x$-axis and it tends to
zero in the steady state. This can further be understood from
Fig.\ \ref{fig:scheme}: after the spin current is excited by
the skew scattering, the spin polarization distributes ($\odot$ and $\otimes$)
anti-symmetrically at $\pm k_y$. However the $y$ components of the 
EMF due to the SOC
(the blue arrows) have the same symmetry and rotate the spin
polarizations to the $-x$ direction (the red dashed arrows). Therefore, the spin
Hall current is converted to the inplane spin polarization. 

\begin{figure}[htp]
\begin{center}\includegraphics[height=6.5cm]{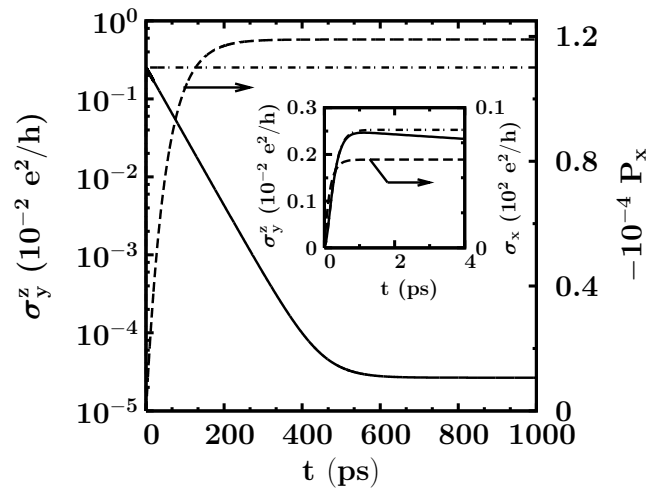}\end{center}
\caption{Time evolution of the SHC with (solid
  curve)/without (chain curve) coherent terms and 
the time evolution of the spin
  polarization (dashed curve) along the $x$-axis at $T=200$\ K.
 $N_i=N_e=4\times 10^{11}$\ cm$^{-2}$ and the electric 
field $E=0.1$\ kV/cm. The corresponding time evolutions of the SHC  
of the first 4\ ps are shown in the inset, together with the 
 charge conductivity $\sigma_x$.
It is noted that the scales of the
 spin polarization/charge conductivity are 
on the right hand side of the frames.}
\label{fig:evolution}
\end{figure}

Now we show the time evolution  of the SHC $\sigma_y^z$ and the spin
polarization $P_x=S^x/N_e$ calculated by numerically solving the KSBE 
with all the scattering explicitly included  at $T=200$\ K 
in Fig.\ \ref{fig:evolution}.
The parameters in the calculation are taken as following:
the electron and impurity densities $N_e=N_i=4\times 10^{11}$\ cm$^{-2}$;  
 $a=7.5$\ nm;  $E=0.1$\ kV/cm;  $\gamma=11.4$\  eV$\cdot$\AA$^3$; and
$\sqrt{\lambda}\gamma=2.07$\ \AA.
From the figure, the SHC  increases
with time first from zero value to a maximum one
in nearly 1\ ps, then decreases slowly to a very small value
(in stead of zero due to the inclusion of the 
Dresselhaus SOC with the cubic ${\bf k}$ terms)  in a
characteristic time scale about 50\ ps. 
The spin polarization is along
the $-x$-axis and increases from zero
to its steady state $P_x=1.2\times10^{-4}$.

The above evolution is easy to understand with the help of the schematic 
in Fig.\ \ref{fig:scheme}. When the positive electric 
field is applied, the skew
scattering scatters the spin-down electrons to the upper panel of the
$\mathbf k$-space and the spin-up ones to the lower panel. This leads to
the spin currents flowing along the $y$-direction. The strength of
the skew scattering is determined by the shift of the electron
distribution, so the SHC increases fast to its maximum value
$\sigma_y^z\sim2.5\times10^{-3}\frac{e^2}{h}$ at the
time scale of the charge current (see the dashed curve in the
inset of Fig.\ \ref{fig:evolution}). 
Then the spin polarization precesses along the $y$-components
of the EMF from the SOC to 
the $-x$ direction to generate the in-plane spin
polarization.  Due to the symmetry, the spin
polarizations along the $y$ and $z$ axes are zero.

It is interesting to see that the obtained SHC in the steady state is 
orders of magnitude smaller than the 
one obtained from the Kubo formula widely used in the
literature where only the lowest
order diagram is considered.\cite{tse06-a}
With only the impurity
scattering included at zero temperature, the Kube formula gives
the extrinsic SHC\cite{tse06-a} as  $\frac{2\pi  m^{\ast}\lambda
  \gamma^2\varepsilon_F}{\hbar^2}\sigma_x\sim 3.37\times
10^{-3} \sigma_x$ with the charge conductivity $\sigma_x=\frac{ne^2\tau}{m}$.
This value is  orders of magnitude larger than our result
 $\sigma_y^z\sim4.1\times10^{-8}\sigma_x$, obtained from the fully microscopic 
KSBE. The difference is caused by the inhomogeneous broadening\cite{mwu}
in spin precession due to the Dresselhaus/Rashba SOC. 
 To show this effect, we
drop the coherent term 
and plot the time evolution of the SHC by 
chain curves in Fig.\ \ref{fig:evolution}. We find that the steady SHC
$\sigma_y^x\sim 0.4 \times
10^{-3}\sigma_x$  is much closer to
the one above given by the Kubo formula at 0\ K. Therefore, we conclude that 
it is not suitable to calculate the extrinsic SHC
without considering the spin precession.
Furthermore, the side jump mechanism gives an
time-independent contribution as
$\sigma_y^z=-2\lambda\gamma^2e^2N_e\sim-2\times10^{-3}{e^2}/{h}$ with
the opposite sign of
the one from the skew scattering.

It is further noted that although the  spin precession induced by the EMF 
leads to the vanishment of the SHC, it 
plays a very important role in generating the spin
polarization. From the discussion above, it is easy to see that the skew
scattering alone cannot generate any spin polarization due to the
anti-symmetrical spin polarization at $\pm k_y$. Differing from the
CISP\cite{hans07} where the SOC is used to provide a
current induced EMF,\cite{mq,kato.sc04} here the SOC  acts
as an anti-symmetrical precession field, 
which is the same as the spin polarization induced by the
spin-dependent phonon scattering.\cite{tarasenko05} 
We further note that the spin polarization induced by the skew scattering is
very different from the CISP:
(i) The spin polarization induced by current induced EMF 
 prefers the small well width which gives
a large EMF,\cite{hans07} whereas the one induced by the skew
scattering and the spin precession prefers a large 
well width. The later can be seen as following:
from Fig.\ \ref{fig:evolution}, the evolution of the
extrinsic spin Hall concurrent can be written as
 $J_y^z(t)=J_{y,0}^ze^{-t/\tau_s}$
with the relaxation time $\tau_s$. $J_{y,0}^z$ approximates to the
maximum value which is only determined by the skew scattering.
 Therefore $S^x={m^{\ast}\tau_s\gamma((\pi/a)^2-\langle
  k_x^2\rangle)}J_{y,0}^z/e$. For the D'yakonov-Perel'
mechanism,\cite{dp.me}
$\tau_s\propto{[\gamma((\pi/a)^2-\langle k_x^2\rangle)]^{-2}}$.
Hence $S^x\propto {\gamma((\pi/a)^2-\langle k_x^2\rangle)}^{-1}$.
(ii) When the electric field is along the $x$-axis, for 
the Dresselhaus SOC,  the CISP is along the
$x$ direction, while the polarization
induced by the skew scattering
is along the $-x$ direction.
 However, for the
strain-induced or the Rashba SOC, the spin polarizations from both mechanisms
are along the same direction.
(iii) The  CISP decreases with the impurity
density\cite{hans07} because high
impurity density reduces the EMF effectively. However, the 
skew-scattering-induced spin polarization
increases with impurity density as the skew scattering is 
proportional to the impurity density.

In summary, we investigate the SHC and the spin polarization
induced by the $\mathbf k$-asymmetric spin-conserving skew scattering 
in symmetrical $(001)$ GaAs quantum well from the
fully microscopic KSBE approach at high temperature with all the
scattering explicitly included. 
We find the spin precession induced by the Dresselhaus/Rashba SOC
has a very important effect on the extrinsic 
SHC and verify the vanishment of the SHC for linear $\mathbf
  k$-dependent SOC.
We also show that the SHC induced
by the skew scattering calculated from the Kubo formula in the
literature
is inadequate without considering the spin precession.
Finally we show that with the joint effects from the skew
scattering and the spin precession, an in-plane 
spin polarization can be generated
which  can be further rotated to the $z$-direction by applying an 
external in-plane magnetic field.

The authors acknowledge valuable discussions with Z. Y. Weng.
This work was supported by the Natural Science Foundation of China
under Grant Nos.\ 10574120 and 10725417, the National Basic
Research Program of China under Grant No.\ 2006CB922005 and
the Knowledge Innovation Project of Chinese Academy of
Sciences.


\begin{thebibliography}{0}
\bibitem{spintronics} {\em Semiconductor Spintronics and Quantum
    Computation}, eds. D. D. Awschalom, D. Loss, and N. Samarth
  (Springer-Verlag, Berlin, 2002);
 I. \v{Z}uti\'{c}, J. Fabian, and  S. Das Sarma,
  Rev. Mod. Phys. {\bf 76}, 323 (2004).
\bibitem{meier} {\it Optical Orientation}, eds. F. Meier and
  B. P. Zakharchenya (North-Holland, Amsterdam, 1984).
\bibitem{tarasenko05}S. A. Tarasenko, Phys. Rev. B {\bf 72}, 113302
  (2005).
\bibitem{ganlchev}S. D. Ganichev, E. L. Ivchenko, V. V. Bel'kov,
  S. A. Tarasenko, M. Sollinger, D. Weiss, W. Wegscheider, and
  W. Prettl, Nature {\bf 417},153 (2002).
\bibitem{ohno}Y. Ohno, D. K. Young, B. Beschoten, F. Matsukura,
  H. Ohno, and D. D. Awschalom, Nature {\bf 402}, 790 (1999).
\bibitem{hans06}H.-A. Engel, E. I. Rashba, and B.
  I. Halperin, cond-mat/0603306.
\bibitem{zsf}S. Zhang, Phys. Rev. Lett. {\bf 85}, 393 (2000).
\bibitem{tse05}W. K. Tse, J. Fabian, I. \v{Z}uti\'c, and S. Das
  Sarma, Phys. Rev. B {\bf 72}, 241303(R) (2005).
\bibitem{nomura} K. Nomura, J. Wunderlich, J. Sinova, B. Kaestner,
  A. H. MacDonald, and T. Jungwirth, Phys. Rev. B {\bf 72}, 245330
  (2005).
\bibitem{kato.sc04}Y. K. Kato, R. C. Myers, A. C. Gossard, and
  D. D. Awschalom, Science {\bf 306}, 1910 (2004).
\bibitem{sih.na05}V. Sih, R. C. Myers, Y. K. Kato, W. H. Lau,
  A. C. Gossard, and D. D. Awschalom, Nature Phys. {\bf 1}, 31
  (2005).
\bibitem{sih.prl06}V. Sih, W. H. Lau, R. C. Myers, V. R. Horowitz,
  A. C. Gossard, and D. D. Awschalom, Phys. Rev. Lett. {\bf 97},
  096605 (2006).
\bibitem{stern.prl06}N. P. Stern, S. Ghosh, G. Xiang, M. Zhu, N. Samarth, and
  D. D. Awschalom, Phys. Rev. Lett. {\bf 97}, 126603 (2006).
\bibitem{wunderlich.prl05}J. Wunderlich, B. Kaestner, J. Sinova, and
  T. Jungwirth, Phys. Rev. Lett. {\bf 94}, 047204 (2005).
\bibitem{ganichev.prl01}S. D. Ganichev, E. L. Ivchenko, S. N. Danilov,
  J. Eroms, W. Wegscheider, D. Weiss, and W. Prettl,
  Phys. Rev. Lett. {\bf 86}, 4358 (2001).
\bibitem{ganichev.na06}S. D. Ganichev, V. V. Bel'kov, S. A. Tarasenko,
  S. N. Danilov, S. Giglberger, C. Hoffmann, E. L. Ivchenko,
  D. Weiss, W. Wegscheider, C. Gerl, D. Schuh, J. Stahl,
  J. D. Boeck, G. Borghs, and W. Prettl, Nature Phys. {\bf 2}, 1609 (2006).
\bibitem{ganichev.cn06}S. D. Ganichev, S. N. Danilov, V. V. Bel'kov,
  S. Giglberger, S. A. Tarasenko, E. L. Ivchenko, D. Weiss,
  W. Jantsch, F. Sch\"affler, D. Gruber, and W. Prettl,
  cond-mat/0610736.
\bibitem{valenzuela.na06}S. O. Valenzuela and M. Tinkham, Nature {\bf
    442}, 176 (2006).
\bibitem{murakami03}S. Murakami, N. Nagaosa, and S. C. Zhang, Science
  {\bf 301}, 1348 (2003).
\bibitem{sinova04}J. Sinova, D. Culcer, Q. Niu, N. A. Sinitsyn,
  T. Jungwirth, and A. H. MacDonald, Phys. Rev. Lett. {\bf 92}, 126603
  (2004).
\bibitem{schliemann06}J. Schliemann, Int. J. Mod. Phys. B {\bf 20},
  1015 (2006).
\bibitem{mishchenko}E. G. Mishchenko, A. V. Shytov, and
  B. I. Halperin, Phys. Rev. Lett. {\bf 93}, 226602 (2004).
\bibitem{lsy06a}S. Y. Liu, X. L. Lei, and N. J. M. Horing,
  Phys. Rev. B {\bf 73}, 035323 (2006);
\bibitem{chen05}W. Q. Chen, Z. Y. Weng, and D. N. Sheng, Phys. Rev. Lett.
  {\bf 95}, 086605 (2005); Phys. Rev. B  {\bf 72}, 235315 (2005).
\bibitem{sheng05}D. N. Sheng, L. Sheng, Z. Y. Weng, and
  F. D. M. Haldane, Phys. Rev. B {\bf 72}, 153307 (2005).
\bibitem{alexander}A. Khaetshii, Phys. Rev. B {\bf 73}, 115323
  (2006).
\bibitem{shytov}A. V. Shytov, E. G. Mishchenko, H.-A. Engel, and
  B. I. Halperin, Phys. Rev. B {\bf 73}, 075316 (2006).
\bibitem{dp.ss} M. I. D'yakonov and V. I. Perel', Zh. Eksp.
 Teor. Fiz. Pis'ma Red. {\bf 13}, 657 (1971)[JETP Lett. {\bf 13},
467 (1971)]; Phys. Lett. {\bf 35A},  459 (1971).
\bibitem{hirsch}J. E. Hirsch, Phys. Rev. Lett. {\bf 83}, 1834 (1999).
\bibitem{tse06-a}W. K. Tse and S. Das Sarma, Phys. Rev. Lett. {\bf
    96}, 056601 (2006).
\bibitem{tse06-b}W. K. Tse and S. Das Sarma, Phys. Rev. B {\bf
      74}, 245309 (2006).
\bibitem{sipe}E. Ya. Sherman, A. Najmaie, H. M. van Driel,
  A. L. Smirl, and J. E. Sipe, Solid State Commun. {\bf 139}, 439
  (2006).
\bibitem{kavokin}A. Kavokin, G. Malpuech, and M. Glazov,
  Phys. Rev. Lett. {\bf 95}, 136601 (2005).
\bibitem{lsy06}S. Y. Liu, N. J. M. Horing, and X. L. Lei,
 Phys. Rev. B {\bf 74}, 165316 (2006).
\bibitem{huang}H. C. Huang, O. Voskoboynikov, and C. P. Lee,
  J. Appl. Phys. {\bf 95},  1918 (2004).
\bibitem{hans05}H.-A. Engel, B. I. Halperin,  and E.
  I. Rashba, Phys. Rev. Lett. {\bf 95}, 166605 (2005).
\bibitem{hankiewicz}E. M. Hankiewicz, G. Vignale, and M. E. Flatt\'e,
  Phys. Rev. Lett. {\bf 97}, 266601 (2006).
\bibitem{roksana}R. Golizadeh-Mojarad and S. Datta, cond-mat/0703280.
\bibitem{Dress}G. Dresselhaus, Phys. Rev. {\bf 100}, 580 (1955).
\bibitem{Rashba} Y. A. Bychkov and E. Rashba,
Zh. \'{E}ksp. Teor. Fiz. {\bf 39}, 66 (1984) [Sov. Phys. JETP {\bf
39}, 78 (1984)].
\bibitem{inoue}J. I. Inoue, G. E. W. Bauer, and L. W. Molenkamp,
  Phys. Rev. B {\bf 70}, 041303(R) (2004).
\bibitem{murakami04}S. Murakami, Phys. Rev. B {\bf 69}, 241202(R)
  (2004).
\bibitem{chalaev}O. Chalaev and D. Loss, Phys. Rev. B {\bf 71},
  245318 (2005).
\bibitem{dimitrova}Ol'ga V. Dimitrova, Phys. Rev. B {\bf 71},
    245327 (2005).
\bibitem{lsy0502392}S. Y. Liu and X. L. Lei, cond-mat/0502392.
\bibitem{bernevig0412550}B. A. Bernevig and S. C. Zhang,
  cond-mat/0412550.
\bibitem{weng} Privite communication with Z. Y. Weng.
\bibitem{averkiev}N. S. Averkiev, L. E. Golub, and M. Willander,
  J. Phys.:Condens. Matter {\bf 14}, R271 (2002).
\bibitem{hans07}H.-A. Engel, E. I. Rashba, and B.
  I. Halperin, Phys. Rev. Lett. {\bf 98}, 036602 (2007).
\bibitem{tarasenko06}S. A. Tarasenko, Pis'ma zh.
Eksp. Teor. Fiz. {\bf 84}, 233 (2006)
 [JETP Lett. {\bf 84}, 199 (2006)].
\bibitem{kato.na03} Y. Kato, R. C. Myers, A. C. Gossard, and
  D. D. Awschalom, Nature {\bf 427}, 50 (2003); Phys. Rev. Lett. {\bf
    93}, 176601 (2004).
\bibitem{trushin}M. Trushin and J. Schliemann, Phys. Rev. B 
{\bf 75}, 155323 (2007).
\bibitem{wu} M. W. Wu, M. Q. Weng, and J. L. Cheng, in {\em
Physics, Chemistry and Application of Nanostructures: Reviews and Short Notes to Nanomeeting 2007}, eds. V. E. Borisenko, V. S. Gurin, and
S. V. Gaponenko (World Scientific, Singapore, 2007) pp. 14, 
and references therein.
\bibitem{wu1}M. W. Wu and H. Metiu, Phys. Rev. B {\bf 61}, 2945 (2000);
M. Q. Weng and M. W. Wu, Phys. Rev. B {\bf 68}, 075312 (2003).
\bibitem{haug} H. Haug and A.P. Jauho, \textit{Quantum Kinetics
in Transport and Optics of Semiconductors} (Springer, Berlin, 1996).
\bibitem{gamma} A. G. Aronov, G. E. Pikus, and A. N. Titkov,
Zh. Eksp. Teor. Fiz. {\bf 84}, 1170 (1983) [Sov. Phys. JETP {\bf 57},
 680 (1983)].
\bibitem{zhou}J. Zhou, J. L. Cheng, and M. W. Wu, Phys. Rev. B
{\bf 75}, 045305 (2007).
\bibitem{mq}M. Q. Weng, M. W. Wu, and L. Jiang, Phys. Rev. B {\bf 69},
  245320 (2004).
\bibitem{mwu} M. W. Wu and C. Z. Ning, Eur. Phys. J. B {\bf 18}, 373 (2000);
 M. W. Wu, J. Phys. Soc. Jpn. {\bf 70}, 2195 (2001).
\bibitem{dp.me}M. I. D'yakonov and V. I. Perel', Zh. Eksp. Teor. Fiz. {\bf
    60}, 1954 (1971) [Sov. Phys. JETP {\bf 38}, 1053 (1971)].
\end{thebibliography}
\end{document}